# TESTING STABILITY AND ROBUSTNESS IN THREE CRYPTOGRAPHIC CHAOTIC SYSTEMS

N. A. Anagnostopoulos[a], K. Konstantinidis[a], A. N. Miliou[a,*]
and S. G. Stavrinides[b]

[a] Department of Informatics, Aristotle University of Thessaloniki, GR 54124, Thessaloniki, Greece
[b] Physics Department, Aristotle University of Thessaloniki, GR 54124, Thessaloniki, Greece

March 5, 2010

In practical applications, it is crucial that the drive-response systems, although identical in all respects, are synchronized at all times, even if there is noise present. In this work, we test the stability and robustness of three distinct and well-known cryptographic chaotic systems, and compare the results in relation to the desired security.

## 1 Introduction

The use of synchronized chaotic systems for communications usually relies on the robustness of the synchronization within the transmitter and receiver pair [1-6]. However, if the communication channel is noisy and/or there is internal noise at the electronic circuitry, the distorted signal at the receiver input might cause considerable synchronization mismatch between the transmitter–receiver pair [7-11].

In this paper, we consider three dynamical systems; the first one is presented in [12], the second one is a Chua-like dynamical system [13], and the third one is a Lorenz-like system presented in [14], and we compare the effect of noise on the synchronization of the systems with different noise levels.

The paper is organized as follows: the circuits' descriptions and the synchronization properties are presented in Section 2. The simulation results obtained are shown in Section 3. Finally, concluding remarks are given in Section 4.

---

[*] Corresponding author. Tel.: +30 2310 998407; fax: +30 2310 998419.
E-mail address: amiliou@csd.auth.gr (A.N. Miliou)

## 2 Description of the Circuits

We present three non-linear cryptographic chaotic circuits and investigate their synchronization's stability and robustness to external noise.

### *2.1 Non-Autonomous, Non-Linear Electronic Circuit*

The first circuit [12] is a combination of two stages (both for the transmitter and the receiver). The first stage is a linear periodic oscillator consisting of an inverting amplifier and two integrators. The second stage that acts as non-linear feedback consists of a comparator, a sub-stage that adjusts voltage levels from $\pm V_{sat}$ Volts to 0 and 5 Volts respectively, an XOR gate, and a buffer with a voltage divider as input. This stage is used to externally trigger the oscillator. The entire circuit produces chaotic oscillations under certain triggering conditions. All amplifiers, integrators, the comparator, and the buffer were implemented with operational amplifiers.

The chaotic voltage across the last of the two integrators is selected and compared with a certain DC voltage level. The result of the comparison is a voltage signal switching between $\pm V_{sat}$ voltage levels, where $\pm V_{sat}$ are the positive and negative saturation levels of the operational amplifier. This signal, after being normalized to 0 and 5 Volts, is multiplexed at the transmitter's XOR gate with the message and transmitted to the receiver. The transmitted signal is used at the receiver to drive the receiver's XOR gate, which decrypts the original message. The transmitter-receiver system (shown in Fig. 1) is governed by the following set of equations:

$$H^S = \begin{cases} -V_{sat}, & V_2 > V_o \\ +V_{sat}, & V_2 < V_o \end{cases} \Leftrightarrow H^S = \begin{cases} 0, & (V_2/V_o) > 1 \\ 1, & (V_2/V_o) < 1 \end{cases} \quad (1)$$

$$RC \frac{dV_2}{dt} = -V_1 - \frac{R}{R_S} V_2 \quad (2)$$

$$RC \frac{dV_1}{dt} = -V - \frac{R}{R_S} V_1 \quad (3)$$

Where $V = -\frac{R}{R_f} V_{out} - V_2$, $V_{out} = U^* \cdot F$, $F = M \oplus H^S$ and $U^* = \kappa \cdot 5\,\text{V}$,

where $\kappa$ is the ratio of potentiometer $R_2$. When we are adding the filter after $V_{out\,2}$

$$V_C = \frac{1}{\sqrt{1 + \omega^2 R_e^2 C_e^2}} V_{out2} \quad (4)$$

$$V_M = \begin{cases} 0\,V, & V_C \leq 2.5\,V \\ 5\,V, & V_C > 2.5\,V \end{cases} \Leftrightarrow V_M = \begin{cases} 0, & V_C \leq 2.5\,V \\ 1, & V_C > 2.5\,V \end{cases} \quad (5)$$

The values of the parameters used in circuit A are given in Table 1.

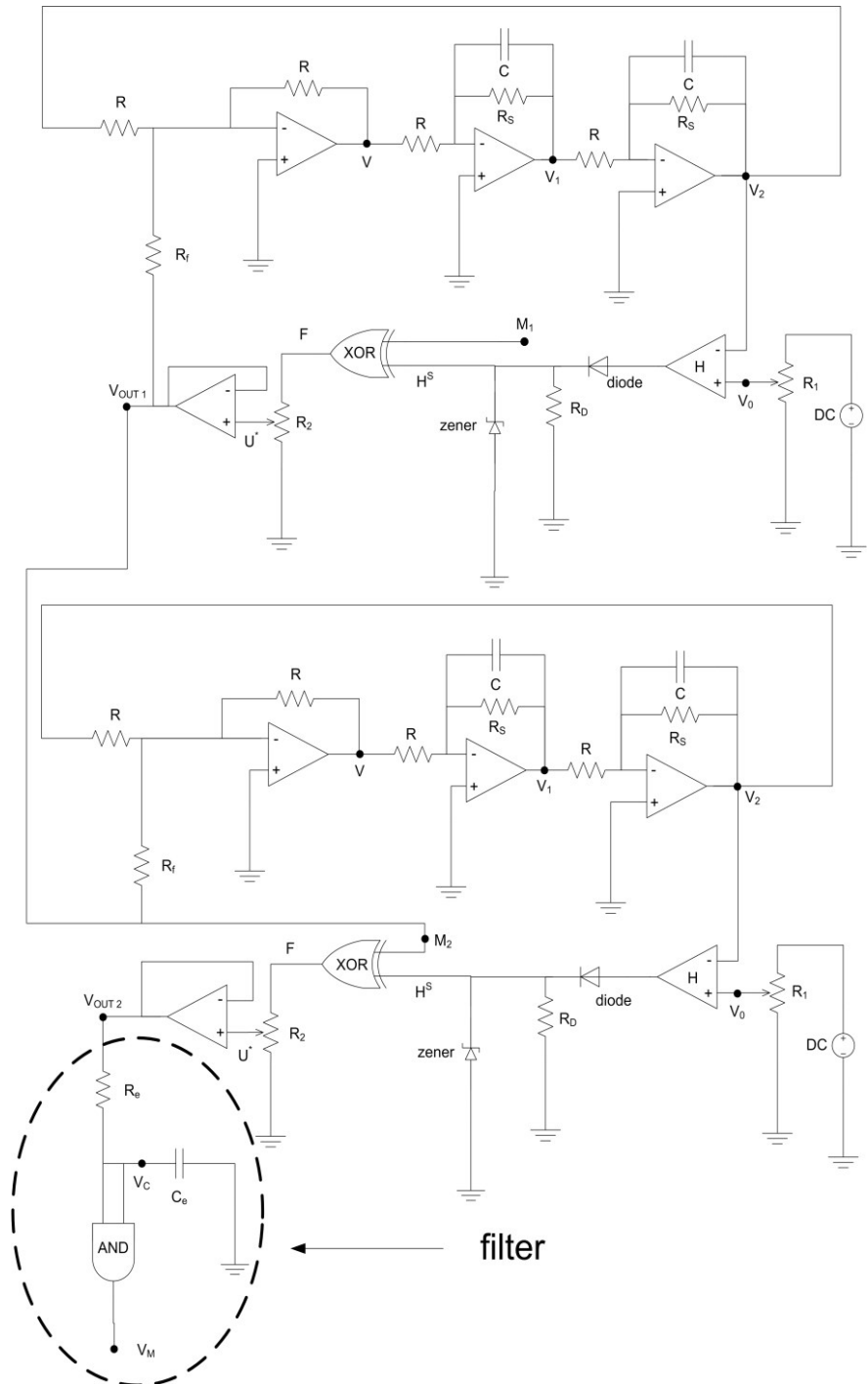

Fig. 1 Schematic diagram of the transmitter-receiver system (circuit A).

Table 1 Parameters' values and types of the elements used in the circuit A

| R | 10 kΩ |
|---|---|
| $R_s$ | 510 kΩ |
| $R_f$ | 18 kΩ |
| $R_1$ | 10 kΩ (potentiometer at 50%) |
| $R_2$ | 10 kΩ (potentiometer at 100%) |
| $R_D$ | 1 kΩ |
| DC voltage | 2 V |
| C | 1 nF |
| diode | 1N4001 |
| zener | 1N4733A |
| AND gate | 74ALS08M |
| XOR gate | 74ALS86M |
| op amp's | TL084CD |

## *2.2 Chua-Like Non-Linear Electronic Circuit*

The second circuit is a combination of two stages (both for the transmitter and the receiver). The first stage is a Chua-like chaotic oscillator [13] where the non-linear resistor (Chua's diode) is implemented with two negative resistors connected in parallel, using two operational amplifiers and six resistors. The circuit has the same functionality as the original Chua's circuit and, with appropriate values of the variable resistor, produces chaotic oscillations.

The second stage consists of a comparator, a sub-stage that adjusts voltage levels from $\pm V_{sat}$ Volts to 0 and 5 Volts respectively, an XOR gate, and a buffer with a voltage divider as input. Both the comparator and the buffer were implemented with operational amplifiers. The receiver is identical to the transmitter except that a low-pass filter is added at the buffer's output, followed by an AND gate.

The chaotic voltage across one of the two capacitors is selected and compared with a certain DC voltage level. The result of the comparison is a voltage signal switching between $\pm V_{sat}$ voltage levels, where $\pm V_{sat}$ are the positive and negative saturation levels of the operational amplifier. This signal, after being normalized to 0 and 5 Volts, is multiplexed at the transmitter's XOR gate with the message and transmitted to the receiver. The transmitted signal is used at the receiver to drive the receiver's XOR gate, which decrypts the original message. There is no feedback, so we use the voltage across the other capacitor as the synchronization signal between transmitter and receiver. The transmitter-receiver system (shown in Fig. 2) is governed by the following set of equations:

$$C_1 \frac{dV_{C1}}{dt} = G(V_{C2} - V_{C1}) - g(V_{C1}) \tag{6}$$

$$C_2 \frac{dV_{C2}}{dt} = G(V_{C1} - V_{C2}) + i_L \tag{7}$$

$$L \frac{di_L}{dt} = -V_{C2} \tag{8}$$

Where $g(V_{C1}) = m_0 V_{C1} + \frac{1}{2}(m_1 - m_0)\left(|V_{C1} + B_p| - |V_{C1} - B_p|\right)$, $G = \frac{1}{R_C}$,

$m_0 = m_{11} + m_{02} = \frac{1}{R_B} - \frac{1}{R_{g1}}$, $\quad m_1 = m_{11} + m_{12} = -\left(\frac{1}{R_{g1}} + \frac{1}{R_{g2}}\right)$ and

$B_p = \frac{R_{g2}}{R_B + R_{g2}} E_{sat}$.

$$V_{out} = U^* \cdot F \tag{9}$$

Where $F = M \oplus H^S$, $H^S = \begin{cases} -V_{saturation}, & V_{C1} > V_o \\ +V_{saturation}, & V_{C1} < V_o \end{cases} \Leftrightarrow H^S = \begin{cases} 0, & (V_{C1}/V_o) > 1 \\ 1, & (V_{C1}/V_o) < 1 \end{cases}$

and $U^* = \kappa \cdot 5\,\text{V}$, where $\kappa$ is the ratio of potentiometer $R_2$.

While when we are adding the filter after $V_{out\,2}$

$$V_C = \frac{1}{\sqrt{1 + \omega^2 R_e^2 C_e^2}} V_{out2} \tag{10}$$

$$V_M = \begin{cases} 0\,V, & V_C \leq 2.5\,V \\ 5\,V, & V_C > 2.5\,V \end{cases} \Leftrightarrow V_M = \begin{cases} 0, & V_C \leq 2.5\,V \\ 1, & V_C > 2.5\,V \end{cases} \tag{11}$$

The values of the parameters used in circuit B are given in Table 2.

Table 2 Parameters' values and types of the elements used in the circuit B

| | |
|---|---|
| $R_A$ | 220 Ω |
| $R_B$ | 22 kΩ |
| $R_C$ | 2 kΩ (potentiometer at 50% [chaotic region ranging from ~2% to ~70%]) |
| $R_{g1}$ | 2.2 kΩ |
| $R_{g2}$ | 3.3 kΩ |
| $R_D$ | 1 kΩ |
| $R_1$ | 10 kΩ (potentiometer at 50%) |
| $R_2$ | 10 kΩ (potentiometer at 100%) |
| $C_1$ | 10 nF |
| $C_2$ | 100 nF |
| L | 18 mH |
| DC voltage | 2 V |
| diode | 1N4001 |
| zener | 1N4733A |
| AND gate | 74ALS08M |
| XOR gate | 74ALS86M |
| op amp's | TL084CD ($E_{sat} \approx 7.5V$ for 9V feed) |

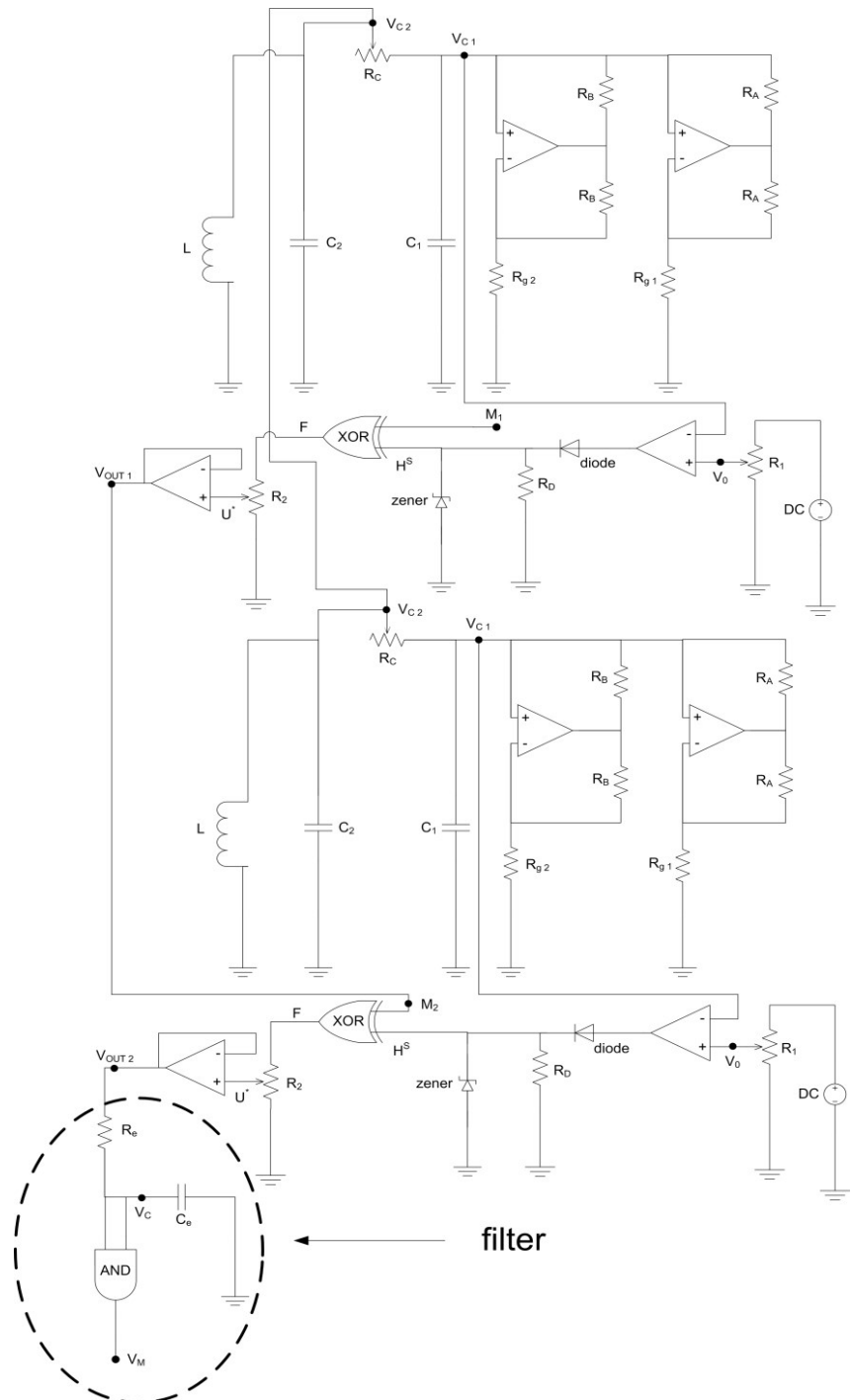

Fig. 2 Schematic diagram of the Chua-like transmitter-receiver system (circuit B).

## 2.3 Lorenz-Like Non-Linear Electronic Circuit

The third circuit is a combination of two stages (both for the transmitter and the receiver). The first stage is a Lorenz-like chaotic oscillator [14]. It consists of three integrators, three summing amplifiers, three inverting amplifiers, two non-inverting amplifiers and two multipliers.

The second stage consists of a comparator, a sub-stage that adjusts voltage levels from $\pm V_{sat}$ Volts to 0 and 5 Volts respectively, an XOR gate, and a buffer with a voltage divider as input. All amplifiers, integrators, the comparator, and the buffer were implemented with operational amplifiers. The receiver is identical to the transmitter except that a low-pass filter is added at the buffer's output, followed by an AND gate.

The chaotic voltage across one of the three capacitors is selected and compared with a certain DC voltage level. The result of the comparison is a voltage signal switching between $\pm V_{sat}$ voltage levels, where $\pm V_{sat}$ are the positive and negative saturation levels of the operational amplifier. This signal, after being normalized to 0 and 5 Volts, is multiplexed at the transmitter's XOR gate with the message and transmitted to the receiver. The transmitted signal is used at the receiver to drive the receiver's XOR gate, which decrypts the original message. Again, there is no feedback, so we use the voltage across one of the other two capacitors as the synchronization signal between the transmitter and the receiver. The transmitter-receiver system (shown in Fig. 3) is governed by the following set of equations:

$$\frac{dV_1}{dt} = -\frac{R_{50}}{R_x CR} V_1 - \frac{1}{R_x CR} V_3 \tag{12}$$

$$\frac{dV_3}{dt} = -\frac{R_{200}}{R_x CR} V_1 - \frac{1}{R_x CR_{200}} \left(1 + \frac{R_{30}}{R_x}\right) V_1 V_2 \tag{13}$$

$$\frac{dV_2}{dt} = -\frac{R_6}{R_x CR_3} V_2 + \frac{1}{R_x C} \left(1 + \frac{R_{30}}{R_x}\right) V_1 V_3 \tag{14}$$

$$V_{out} = U^* \cdot F \tag{15}$$

where $F = M \oplus H^S$, $H^S = \begin{cases} -V_{saturation}, & V_3 > V_o \\ +V_{saturation}, & V_3 < V_o \end{cases} \Leftrightarrow H^S = \begin{cases} 0, & (V_3/V_o) > 1 \\ 1, & (V_3/V_o) < 1 \end{cases}$

and $U^* = \kappa \cdot 5\,V$, where $\kappa$ is the ratio of potentiometer $R_2$.

While when we are adding the filter after $V_{out\,2}$

$$V_C = \frac{1}{\sqrt{1 + \omega^2 R_e^2 C_e^2}} V_{out2} \tag{16}$$

$$V_M = \begin{cases} 0\,V, & V_C \leq 2.5\,V \\ 5\,V, & V_C > 2.5\,V \end{cases} \Leftrightarrow V_M = \begin{cases} 0, & V_C \leq 2.5\,V \\ 1, & V_C > 2.5\,V \end{cases} \tag{17}$$

The values of the parameters used in circuit C are given in Table 3.

Fig. 3 Schematic diagram of the Lorenz-like transmitter-receiver system (circuit C).

Table 3 Parameters' values and types of the elements used in the circuit C

| | |
|---|---|
| R | 10 kΩ |
| $R_{50}$ | 50 kΩ |
| $R_x$ | 1 kΩ |
| $R_{200}$ | 200 kΩ |
| $R_y$ | 2 kΩ |
| $R_{30}$ | 30 kΩ |
| $R_3$ | 3 kΩ |
| $R_6$ | 6 kΩ |
| $R_{100M}$ | 100 MΩ |
| $R_1$ | 10 kΩ (potentiometer at 50%) |
| $R_2$ | 10 kΩ (potentiometer at 100%) |
| $R_D$ | 1 kΩ |
| C | 100 nF |
| DC1 voltage | 2 V |
| DC2 voltage | -20 mV |
| diode | 1N4001 |
| zener | 1N4733A |
| AND gate | 74ALS08M |
| XOR gate | 74ALS86M |
| op amp's | TL084CD |

## 3  Simulation Results

We have selected Multisim to simulate the three chaotic systems, since it provides an interface similar to the real implementation environment. External noise is applied on the communication channel, while different noise amplitudes *A* have been utilized, ranging from 0.01% to 50% of the mean signal amplitude. As the noise amplitude *A* is increased, the synchronization of the system continuously deteriorates and is practically destroyed in all cases above a certain noise level.

A thermal noise generator connected at the communication channel provided artificial external noise added at the simulation. The information signal used is a square pulse with a frequency of 6.222 KHz, ranging from 0 to 5 V, and being connected at the XOR gate of each circuit's transmitter. Both the transmitter and the receiver are identical circuits. The filters used in the simulations are presented in Table 4.

Table 4 Filter parameters

| | |
|---|---|
| $R_e$ | Filter 1 $R_e$=40 Ω<br>Filter 2 $R_e$=1 kΩ<br>Filter 3 $R_e$=1 kΩ |
| $C_e$ | Filter 1 $C_e$=7 nF<br>Filter 2 $C_e$=2.5 nF<br>Filter 3 $C_e$=7 nF |

### 3.1 Circuit A

Figs. 4 and 5 show the response of circuit A, respectively, without noise and with the application of 10% noise on the communication channel, which also acts as the synchronization channel. Different noise amplitudes $A$ have been utilized and the synchronization of the system deteriorates severely above a 10% noise level.

It is apparent that the signal at the receiver without filtering shows some glitches due to the response of the circuit's elements, which disappear when a filter is added at the receiver's output. In Fig. 5, filter 2 eliminates the glitches completely compared to filter 1, although both filters introduced a slight time shift in the response signal.

However, the synchronization is good (up to the 10% noise level), and the transmitted information is received and reconstructed (decoded) at the output with good quality.

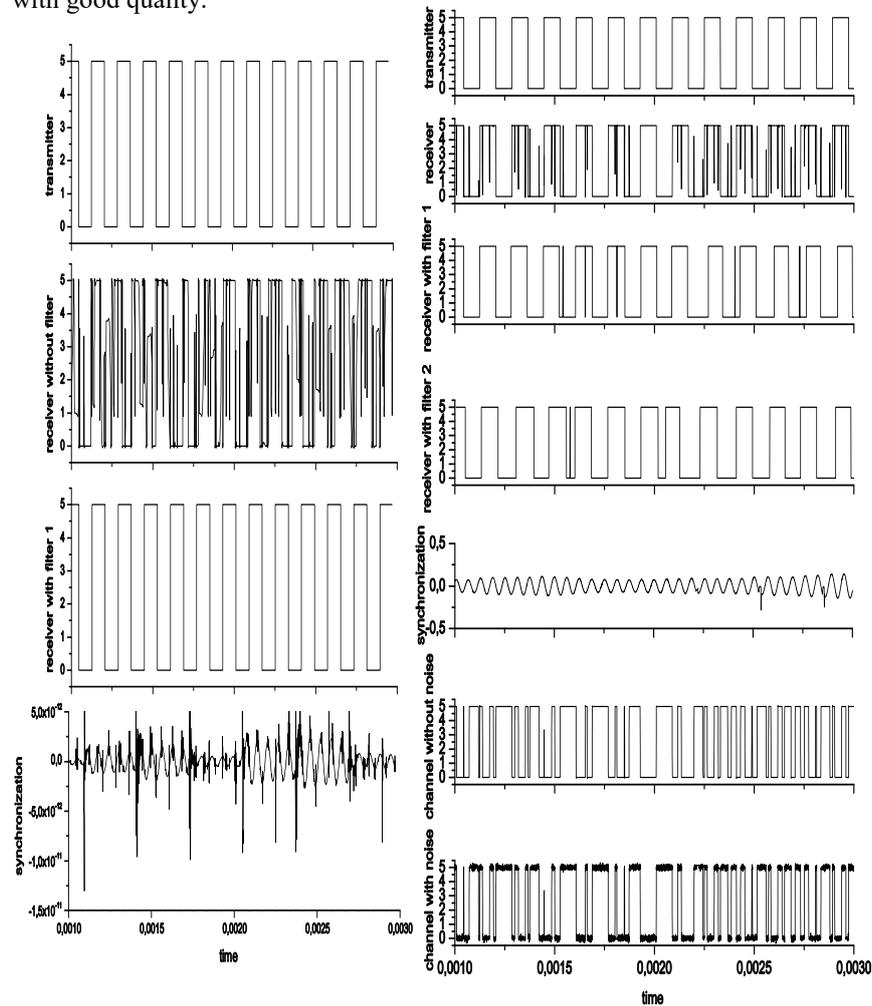

Fig. 4 Response of circuit A without noise.  Fig. 5 Response of circuit A with 10% noise.

### 3.2 Circuit B (Chua-like)

In circuit B, we used two communication channels, one for synchronization and one for information transmission. All simulations of circuit B at the receiver are performed with the use of Filter 1.

Similarly with the previous circuit, we experimented with different noise levels and the synchronization is very good up to the 10% noise level, as shown in Figs. 6 and 7, respectively. The noise was added either to the synchronization channel only or to the information channel only or to both channels. However, one could observe that in the third case, at the receiver with noise in both channels, the response signal demonstrates a transient behavior before perfect synchronization is achieved, as shown in Fig. 7.

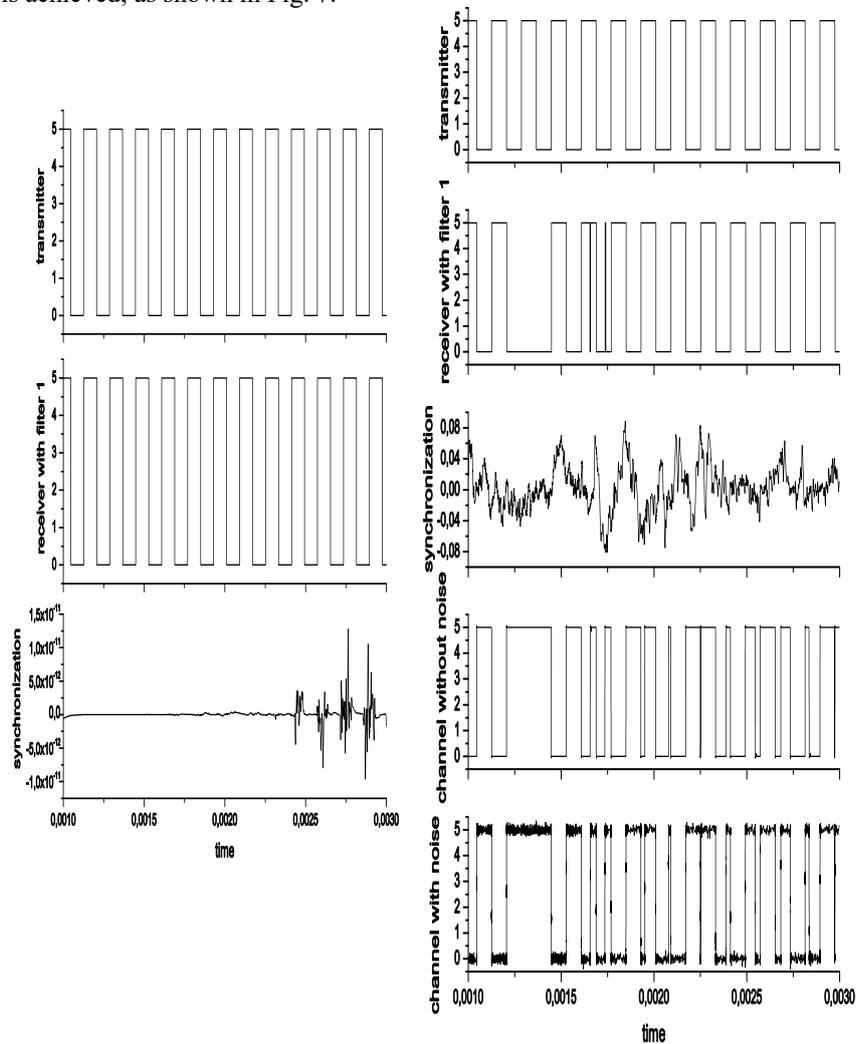

Fig. 6 Response of circuit B without noise.   Fig. 7 Response of circuit B with 10% noise.

### 3.3 Circuit C (Lorenz-like)

In circuit C, as in B, we used two communication channels, one for synchronization and one for information transmission, while Filter 3 is used at the receiver.

Additionally, in this system, we can distinguish two separate cases (Ca and Cb); the first one where we use $V_1$ as the synchronization voltage, and $V_3$ is the voltage that goes to the comparator (Figs. 8 and 9). The schematic diagram in Fig. 3 corresponds to this case. In the second case, we are using $V_3$ as the synchronization voltage, and $V_1$ as the voltage that goes to the comparator (Figs. 10 and 11).

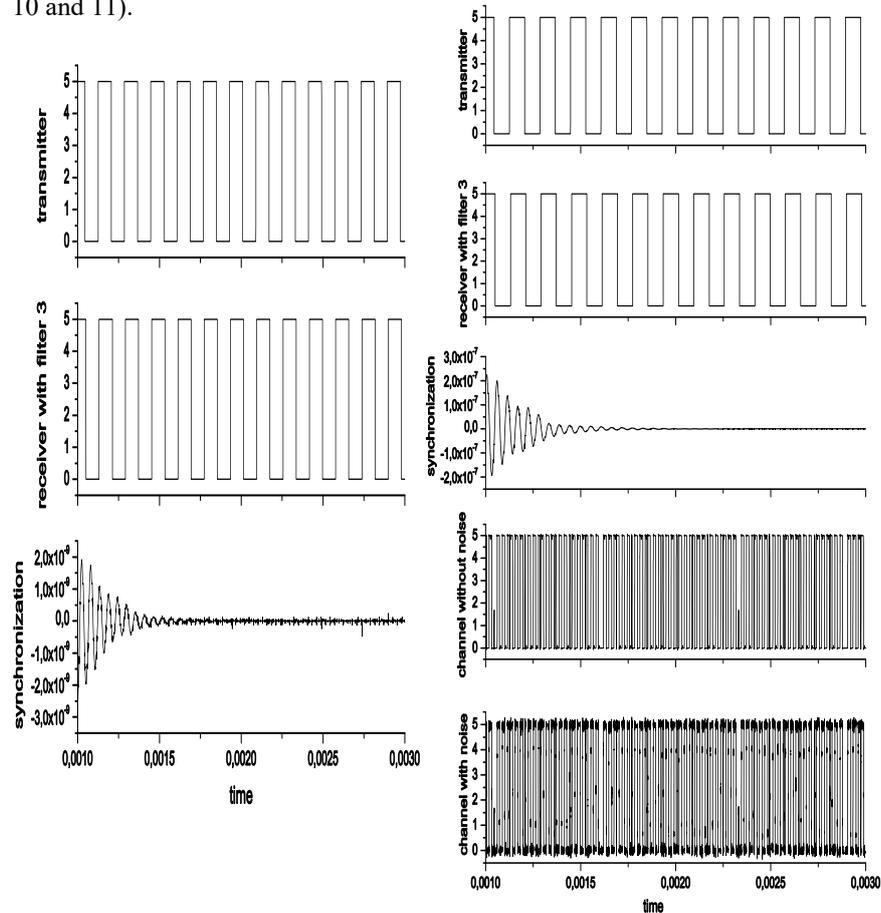

Fig. 8 Response of circuit Ca without noise.  Fig. 9 Response of circuit Ca with 1% noise.

Figs. 8 and 9 show the response of circuit Ca without noise and with the application of 1% noise on the communication (synchronization and information) channels, respectively. Different noise amplitudes $A$ have been utilized, and the synchronization of the system deteriorates severely above the 1% noise level. However, it is apparent that the signal at the receiver with noise in both the information and the synchronization channels is decoded with no impairment at the present level of noise. The synchronization of the transmitter–receiver system is very good up to the 1% noise level.

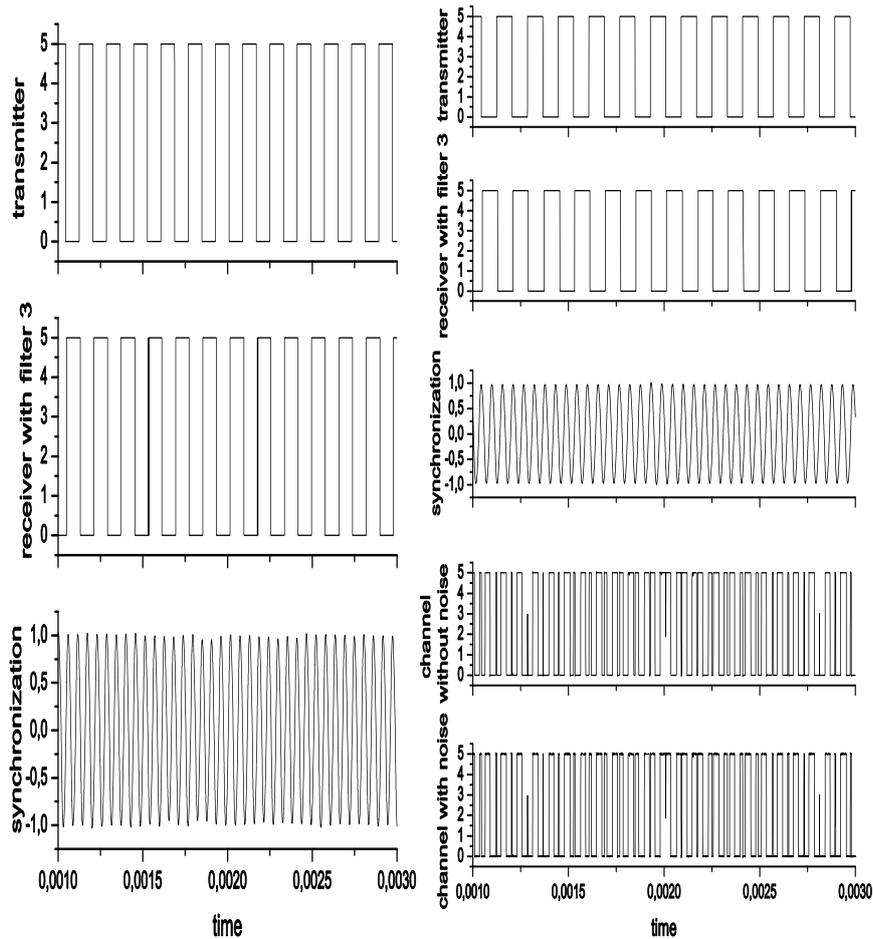

Fig. 10 Response of circuit Cb w/o noise.   Fig. 11 Response of circuit Cb with 1% noise.

Figs. 10 and 11 depict the response of circuit Cb without and with noise, respectively. In this case, we observed an anti-synchronization phenomenon in the coupled Lorenz-like chaotic oscillators depending on the initial conditions [15]. Anti-synchronization has potential applications in practical fields such as secure communications and digital transmission. As in circuit Ca, the system is robust up to the 1% noise level, and the received signal although anti-synchronized is decoded with no impairments.

## 4 Concluding remarks

The use of synchronized chaotic systems for communications relies on the robustness of the synchronization to perturbations in the drive signal.

In this work we have studied the influence of noise on the synchronization of the transmitter–receiver pairs based on the nonlinear circuits presented in [12, 13, 14] by simulating the behavior of the systems in Multisim. The results have demonstrated the robustness of these systems with regards to their synchronization for similar levels of noise amplitudes. Specifically, circuits A and B are withstanding higher amplitude noise perturbations in comparison to circuit C. Among circuits A and B, the latter one uses a filter with a lower cut-off frequency in order to eliminate the glitches appearing at the receiver and, even though it shows some transient behavior, its output is of better quality than A. Although circuit C uses the filter with the lowest cut-off frequency, it decodes the signal without flaws even in the anti-synchronization (Cb) case.

Concluding our experimental study, we could state that there is no doubt that robustness to noise is very advantageous for the synchronization of a system, especially when the system is realized with off-the-shelf electronics.

**Acknowledgements**

This work was supported by NATO ICS.EAP.CLG 983334.